\documentclass[12pt]{article}
\usepackage{amsmath,amssymb,bm,graphicx}
\usepackage{color} 
\usepackage{hyperref}
\usepackage{cite}	

\begin{document}

\begin{flushright}
\end{flushright}

\vskip 0.5 truecm

\begin{center}
{\Large{\bf Berry's phase and quantum mechanical formulation of anomalous Hall effect}}
\end{center}
\vskip .5 truecm
\centerline{\bf Kazuo Fujikawa~$^1$
 {\rm and} Koichiro Umetsu~$^2$}
\vskip .4 truecm
\centerline {\it $^1$~Interdisciplinary Theoretical and Mathematical Sciences Program (iTHEMS), 
}
\centerline {\it RIKEN, Wako 351-0198, 
Japan}
\vskip 0.4 truecm
\centerline {\it $^2$~Laboratory of Physics, College of Science and Technology, and Junior College, }
\centerline{\it Funabashi Campus, Nihon University, Funabashi, Chiba 274-8501, Japan}
\vskip 0.5 truecm
\makeatletter
\makeatother


\begin{abstract}
 The canonical commutation relations in quantum mechanics are not maintained in the anomalous Hall effect described by Berry's phase in the presence of the electromagnetic vector potential. To define quantum mechanical formulation, one may avoid the electromagnetic vector potential but then the anomalous Nernst effect induced by Berry's curvature is not described using a modified phase space volume, although the anomalous Nernst effect by itself may be generated by the adiabatic Berry's curvature.
We also comment on the Born-Oppenheimer approximation if it describes the anomalous Hall effect in the absence of the electromagnetic vector potential in quantum mechanics. An alternative view of the Bjorken-Johnson-Low prescription which is consistent with the principle of quantum mechanics and the existing Berry's phase theory of the anomalous Hall effect is also mentioned.
\end{abstract}


\section{Introduction}
Berry's phase \cite{Berry} is defined as an adiabatic phase in quantum mechanics \cite{Born2} and it is well known that Berry's phase provides an effective description of the involved processes in condensed matter physics. For example,
in the case of the anomalous Hall effect, the actual analysis of the physical phenomena is very involved \cite{Karplus, Luttinger} but the general aspects of physics are well described by the adiabatic Berry's phase \cite{Nagaosa1, Sinitsyn, Nagaosa, Niu0, Niu, Fang}. This effective description is often called {\em a semi-classical description} \cite{Sinitsyn} since, for example, the description of the spin degrees of freedom and the effects of impurities are missing. The present paper is interested in this general quantum mechanical aspect of Berry's phase (although Berry's phase itself is known to exist in classical physics also \cite{Pancharatnam}).
If one defines the quantum mechanical formulation on the basis of the adiabatic effective Lagrangian, the non-canonical commutation relations, which are not quantum mechanical, are inevitably induced in the presence of the electromagnetic vector potential together with adiabatic Berry's connection \cite{Niu2, Duval}. 
To define a quantum mechanical formulation, one may avoid the electromagnetic vector potential. In the absence of the electromagnetic vector potential, however, we lose a means to identify the source current of electromagnetism. As a technical issue, we thus 
failed to show that the induced flow of the electric charge in the so-called gauge invariant formulation of the anomalous Hall effect is precisely transverse to the direction of the applied external force in quantum mechanics. Also we comment on the anomalous Hall effect based on the Born-Oppenheimer approximation \cite{Born, Blount} by making a diagonal dominance approximation from a different point of view. Both Berry's phase and the Born-Oppenheimer approximations fail in the presence of the electromagnetic vector potential in quantum mechanics. 
 
 The deformed Liouville measure has also been used to describe the modification of the electron density of states in the anomalous Hall effect \cite{Niu2, Duval}. In a quantum mechanical framework, however, Berry's connection and the electromagnetic vector potential are not incorporated simultaneously. Thus a modification in the Liouville volume element $dV(1+e\vec{B}\cdot \vec{\Omega})$ with $\vec{\Omega}$ standing for Berry's curvature, which was used to describe the anomalous Nernst effect induced by Berry's curvature \cite{Sinitsyn}, is absent if one sets $\vec{B}=0$, although the anomalous Nernst effect itself may be induced by the adiabatic Berry's curvature\cite{Luttinger2}.

 In a purely quantum mechanical formulation of Berry's phase theory of the anomalous Hall effect, many of the commonly accepted aspects of the adiabatic approximation thus need to be re-examined. Our main motivation to clarify these subtle quantum mechanical aspects is to understand better the recent discussions of the possible connection between Berry's phase and quantum anomalies \cite{Fujikawa2} which are purely quantum mechanical phenomena. An alternative view of the Bjorken-Johnson-Low prescription is also explained, where the quantum mechanical origin of Berry's phase is made transparent and most of the known properties of the existing Berry's phase theory of the anomalous Hall effect are argued to be consistently understood in the framework of quantum mechanics. Some of general aspects of these issues were discussed elsewhere \cite{Fujikawa2,Fujikawa1}; we here present more detailed accounts. 

\section{Berry's phase}
Berry's phase carries its own $U(1)$ gauge symmetry, but the gauge symmetry of Berry's phase and the electromagnetic $U(1)$ symmetry are shown to be incompatible in quantum mechanics. It is known that the semi-classical action of the anomalous Hall effect constrained to a single band is neatly summarized by \cite{Niu2, Duval, Faddeev} 
\begin{eqnarray}\label{action}
S=\int dt\left[ p_{k}\dot{x}_{k} - eA_{k}(\vec{x})\dot{x}_{k} + {\cal A}_{k}(\vec{p})\dot{p}_{k}-\epsilon_{n}(\vec{p})+e\phi(\vec{x})\right].
\end{eqnarray}
We use the notation of the momentum $\vec{p}=\hbar \vec{k}$ rather than the more common wave vector $\vec{k}$ in condensed matter physics. Instead, we use $k, l$ and $m$ mainly as the dummy three-vector indices in the present paper. The four-vector $(e\phi(\vec{x}), eA_{k}(\vec{x})) $ stand for electromagnetic potentials; ${\cal A}_{k}(\vec{p})$ and $\epsilon_{n}(\vec{p})$ are Berry's connection and the ($n$-th band) energy, respectively.
If the action principle is assumed to be valid for \eqref{action}, the equations of motion of the anomalous Hall effect are obtained \cite{Niu0, Nagaosa, Niu}
\begin{eqnarray}\label{effective equations}
\dot{x}_{k}=-\Omega_{kl}(\vec{p})\dot{p}_{l} +\frac{\partial \epsilon_{n}(\vec{p})}{\partial p_{k}}, \ \
\dot{p}_{k}=-eF_{kl}(\vec{x})\dot{x}_{l} + e\frac{\partial}{\partial x_{k}}\phi(\vec{x}).
\end{eqnarray}
The transverse velocity term $-\Omega_{kl}(\vec{p})\dot{p}_{l}$ with $\Omega_{kl}(\vec{p})$ standing for Berry's curvature is responsible for the anomalous Hall effect \cite{Nagaosa1, Sinitsyn, Nagaosa}. These two equations \eqref{effective equations} are formally invariant under respective $U(1)$ gauge symmetries; the electromagnetic $U(1)$ symmetry and the $U(1)$ symmetry of Berry's connection. The Lagrangian \eqref{action} is obtained by adding Berry's phase, ${\cal A}^{(n)}_{k}(\vec{p})\dot{p}_{k}\sim {\phi^{(n)}}^{\ast}(\vec{p})i\frac{\partial}{\partial{p_{k}}}\phi^{(n)}(\vec{p})\dot{p}_{k}$ where $\phi^{(n)}(\vec{p})$ describes one of the bands \cite{Nagaosa1, Niu}, to the electromagnetic gauge invariant action of a {\em slower} system. But this construction 
is not compatible with the principle of quantum mechanics \footnote{Simply stated, the principle of quantum mechanics is understood here as satisfying the two properties: Characterized by canonical forms of commutation relations and the Hamiltonian is written in terms of covariant derivatives of gauge variables involved.}. (A more detailed proof of this fact will be shown later.) By treating Berry's connection ${\cal A}_{k}(\vec{p})$ as a local object, the above Lagrangian is characterized by the deformed Poisson brackets \cite{Duval, Faddeev}
\begin{eqnarray}\label{modified commutators}
&&\{x_{k},x_{l}\}_{P}=\frac{\epsilon^{klm}\Omega_{m}}{1+e\vec{B}\cdot\vec{\Omega}} , \ \ \ \{p_{k},x_{l}\}_{P}=-\frac{\delta_{kl}+e\Omega_{k}B_{l}}{1+e\vec{B}\cdot\vec{\Omega}},\nonumber\\
&&\{p_{k},p_{l}\}_{P}=- \frac{\epsilon^{klm}eB_{m}}{1+e\vec{B}\cdot\vec{\Omega}},
\end{eqnarray}
where $\epsilon^{klm}B_{m}$ and $\epsilon^{klm}\Omega_{m}$ stand for the curvature tensors defined by $A_{k}(\vec{x})$ and ${\cal A}_{k}(\vec{p})$, respectively. The Poisson brackets here are defined in the conventional manner. The gauge symmetry associated with Berry's connection is defined in the framework of an adiabatic Lagrangian \eqref{action} and the principle of quantum mechanics (canonical commutation relations) is modified when the electromagnetic vector potential is present in addition as in \eqref{modified commutators}. The non-Hamiltonian character of \eqref{action} and \eqref{effective equations} is 
well-recognized in the literature \cite{Niu2,Duval}. It is known that a generalized Lagrangian formalism, which is not quantum mechanical in a conventional sense, consistently accounts for the specific system \eqref{action}, \eqref{effective equations} and \eqref{modified commutators} \cite{Duval, Faddeev}. But Berry's phase is treated as a given object there and the derivation of Berry's phase from another (fast moving) quantum system and the associated restrictions are not transparent. 

Since the derivation of Berry's phase is assumed to be based on the purely quantum mechanical adiabatic process \cite{Berry}, we analyze the above system \eqref{action}, \eqref{effective equations} and \eqref{modified commutators} in the framework of conventional quantum mechanics. We start with a derivation of \eqref{action} in a purely quantum mechanical system with the Hamiltonian of the form, which we believe to be faithful to the original idea of Berry, 
\begin{eqnarray}\label{starting Hamiltonian}
H=H_{0} + H_{1}
\end{eqnarray}
 assuming that the slower particle is charged with $q=-e$ ($e>0$) and the fast particle is neutral, 
\begin{eqnarray}\label{starting Hamiltonian2}
H_{0}(X, P+ eA(X))&=&\frac{1}{2M}(P_{k} + eA_{k}(X))^{2} -e\phi(X),\nonumber\\
H_{1}(x, p; P+ eA(X))&=&H_{1}(x_{k}, p_{k} ; P_{k} + eA_{k}(X)).
\end{eqnarray}
We assume that the canonical quantization of $H$ gives for fast variables
 \begin{eqnarray}\label{standard commutator-0}
 [p_{k}, x_{l}]=\frac{\hbar}{i}\delta_{kl}, \ \ [p_{k}, p_{l}]=0, \ \ [x_{k}, x_{l}]=0
 \end{eqnarray}
 and the canonical quantization of slower variables
 \begin{eqnarray}\label{standard commutator-2}
 [P_{k}, X_{l}]=\frac{\hbar}{i}\delta_{kl}, \ \ [P_{k}, P_{l}]=0, \ \ [X_{k}, X_{l}]=0
 \end{eqnarray}
by treating $H_{0}(X, P + eA(X))$ as the slower system.
Note that the lower case characters stand for the fast variables and the capital characters stand for the slower variables from now on.
We start with the fundamental path integral (the Lagrangian formalism) corresponding to the Hamiltonian \eqref{starting Hamiltonian}
\begin{align}\label{starting path integral}
&Z
=\int {\cal D}\overline{P}_{k}{\cal D}X_{k}\exp \left\{ \frac{i}{\hbar}\int_{0}^{T} dt \left[(\overline{P}_{k}(t)-eA_{k}(X(t)))\dot{X}_{k}(t)
- H_{0}(X(t), \overline{P}(t)) \right] \right\}\notag\\
&\times\int {\cal D}\psi^{\ast}{\cal D}\psi\exp \left\{\frac{i}{\hbar}\int_{0}^{T} dt\int d^{3}x \left[\psi^{\ast}(t,\vec{x})i\hbar \partial_{t}\psi(t,\vec{x}) - \psi^{\ast}(t,\vec{x})H_{1}\left(\frac{\hbar}{i}\vec{\nabla}, \vec{x}; \overline{P}(t) \right)\psi(t,\vec{x}) \right] \right\} 
\end{align}
with the covariant derivative 
\begin{eqnarray}\label{covariant derivative1}
\overline{P}_{k}(t)=P_{k}(t) + eA_{k}(X(t))
\end{eqnarray}
 which is now treated as an independent dynamical variable
using the invariance of the path integral measure
${\cal D}P_{k}{\cal D}X_{k}={\cal D}\overline{P}_{k}{\cal D}X_{k}$. We are operating in the Lagrangian formalism, and the canonical commutation relations are preserved in \eqref{starting path integral} if one goes back to the Hamiltonian formalism, for example, using \eqref{covariant derivative1}
\begin{eqnarray}\label{alternative form of canonical commutators}
[\overline{P}_{k},X_{l}]=-i\hbar\delta_{kl}, \ \ [\overline{P}_{k}, \overline{P}_{l}]=-i\hbar F_{kl}, \ \ [X_{k},X_{l}]=0.
\end{eqnarray}
But the canonical commutation relations are substantially changed at the final stage after applying the adiabatic approximation, as is shown below. We use the second quantized formulation for the fast particle in \eqref{starting path integral} since it is convenient to formulate Berry's phase \cite{Fujikawa2}. We expand the second quantized fast particle in the manner 
\begin{eqnarray}
\psi(t, x)= \sum_{n} a_{n}(t)\phi_{n}(x,\overline{P}(t) )
\end{eqnarray}
with the instantaneous (fixed $\overline{P}(t)$) eigenfunctions
\begin{eqnarray}\label{instantaneous function}
&&H_{1}\left(\frac{\hbar}{i}\vec{\nabla}, \vec{x}; \overline{P}(t) \right)\phi_{n}(x,\overline{P}(t) )=E_{n}(\overline{P}(t))\phi_{n}(x,\overline{P}(t) ), \nonumber\\ 
&&\int d^{3}x\phi^{\ast}_{n}(x,\overline{P}(t) )\phi_{m}(x,\overline{P}(t) )=\delta_{nm}.
\end{eqnarray}
The second quantized part in \eqref{starting path integral} is then written as 
\begin{eqnarray}\label{Auxialy expression}
&&\int \prod_{n}{\cal D}a^{\ast}_{n}{\cal D}a_{n}\exp \left\{ \frac{i}{\hbar}\int_{0}^{T} dt\sum_{n} \left[ a_{n}^{\ast}(t)i\hbar \partial_{t}a_{n}(t)- E_{n}(\overline{P}(t)) a_{n}^{\ast}(t)a_{n}(t) \right] \right\}\nonumber\\
&&\hspace{2cm} \times \exp \left\{\frac{i}{\hbar}\int_{0}^{T} dt \sum_{n,l}\langle n,t|i\hbar\partial_{t}|l,t\rangle a_{n}^{\ast}(t)a_{l}(t) \right\}\nonumber\\
&\simeq&
\int \prod_{n}{\cal D}a^{\ast}_{n}{\cal D}a_{n}\exp \left\{\frac{i}{\hbar}\int_{0}^{T} dt\sum_{n} \left[ a_{n}^{\ast}(t)i\hbar \partial_{t}a_{n}(t)\right] \right\}\nonumber\\
&&\hspace{2cm} \times \exp \left\{-\frac{i}{\hbar}\int_{0}^{T} dt \sum_{n} \left(E_{n}(\overline{P}) -{\cal A}^{(n)}_{k}(\overline{P})\dot{\overline{P}^{k}}\right)a_{n}^{\ast}(t)a_{n}(t)
\right\}
\end{eqnarray}
where we made the {\em adiabatic (diagonal) approximation} in the second expression
\begin{eqnarray}
&&\sum_{n,m}\int_{0}^{T} dt\langle n,t|i\hbar\partial_{t}|m,t\rangle a_{n}^{\ast}(t)a_{m}(t)\nonumber\\
&&=\sum_{n,m}\int_{0}^{T} dt\int d^{3}x\phi^{\ast}_{n}(x,\overline{P}(t) )i\hbar\partial_{t}\phi_{m}(x,\overline{P}(t) )a^{\ast}_{n}(t)a_{m}(t)\nonumber\\
&&\simeq\sum_{n}\int_{0}^{T} dt {\cal A}^{(n)}_{k}(\overline{P})\dot{\overline{P}_{k}}a^{\ast}_{n}(t)a_{n}(t)
\end{eqnarray}
with
\begin{eqnarray}\label{Berry's phase 2}
{\cal A}^{(n)}_{k}(\overline{P})=\int d^{3}x\phi^{\ast}_{n}(x,\overline{P}(t) )i\hbar\partial_{\overline{P}(t)_{k}}\phi_{n}(x,\overline{P}(t) ).
\end{eqnarray}
 Namely, we assume that the off-diagonal parts $\phi^{\ast}_{n}(x,\overline{P}(t) )i\hbar\partial_{t}\phi_{m}(x,\overline{P}(t) )$ with $n\neq m$ are ignored compared to the diagonal parts $\phi^{\ast}_{n}(x,\overline{P}(t) )i\hbar\partial_{t}\phi_{n}(x,\overline{P}(t) )$ by noting that the non-diagonal terms may be ignored if one assumes that all the terms with the slow variable $\dot{\overline{P}}(t)$ are small compared to the energy $E_{n}(\overline{P})$; the off-diagonal terms give a small contribution which is second order in small quantities in the adiabatic approximation. This is an explicit breaking of symmetry, and the quantum coherence among different fast states $\phi_{n}(x,\overline{P}(t) )$ is lost in this approximation. Defining the Schr\"{o}dinger operator expression (i.e., replacing $\hat{a}_{n}(t) \rightarrow \hat{a}_{n}(0)$) in \eqref{Auxialy expression} , we have the evolution operator in the operator formalism \cite{Deguchi}
 \begin{eqnarray}\label{transformation function}
 \langle m|T^{\ast}\exp \left\{-\frac{i}{\hbar}\int_{0}^{T} dt \sum_{l} \left(E_{l}(\overline{P}) -{\cal A}^{(l)}_{k}(\overline{P})\dot{\overline{P}^{k}}\right)\hat{a}_{l}^{\dagger}(0)\hat{a}_{l}(0)
\right\}|n\rangle,
\end{eqnarray}
 with the $n$-th state $|n\rangle=\hat{a}^{\dagger}_{n}(0)|0\rangle$ and $T^{\ast}$ stands for the time ordering operation. Only the diagonal term $m=n$ survives in \eqref{transformation function}, and one has (in an approximate sense as described above)
\begin{eqnarray}
\exp \left\{-\frac{i}{\hbar}\int_{0}^{T} dt \left(E_{n}(\overline{P}) -{\cal A}^{(n)}_{k}(\overline{P})\dot{\overline{P}^{k}}\right)\right\}
\end{eqnarray}
for a fast moving state defined by $|n\rangle=\hat{a}^{\dagger}_{n}(0)|0\rangle$. 
For the periodic motion $\overline{P}(0)=\overline{P}(T)$ this formula is invariant under the gauge transformation 
\begin{eqnarray}\label{gauge transformation2}
{\cal A}^{(n)}_{k}(\overline{P}(t))\rightarrow {\cal A}^{(n)}_{k}(\overline{P}(t)) +\hbar \partial_{k}\alpha_{n}(\overline{P}(t))
\end{eqnarray}
which is induced by $\phi_{n}(x,\overline{P}(t)) \rightarrow \exp[-i\alpha_{n}(\overline{P}(t))]
\phi_{n}(x,\overline{P}(t))
$ if one remembers the definition of Berry's connection.

Adding this final expression to the 
Lagrangian of slower variables appearing in the path integral \eqref{starting path integral} one obtains the Lagrangian defined on the specific state $|n\rangle=\hat{a}^{\dagger}_{n}(0)|0\rangle$ 
of the fast particle
\begin{eqnarray}\label{effective Lagrangian-n}
L_{n}=(\overline{P}_{k}-eA_{k}(X))\dot{X}_{k} +{\cal A}^{(n)}_{k}(\overline{P})\dot{\overline{P}_{k}}
- (\frac{1}{2M}\overline{P}_{k}^{2}+E_{n}(\overline{P})) + e\phi(X)
\end{eqnarray}
which agrees with the common Lagrangian in condensed matter physics \eqref{action} if one identifies
\begin{eqnarray}
\overline{P}_{k}\rightarrow p_{k}, \ \ X_{k}\rightarrow x_{k},\ \ {\rm and}\ \ \epsilon_{n}(p)=\frac{1}{2M}\overline{P}_{k}^{2} + E_{n}(\overline{P}), 
 \end{eqnarray}
 and if one uses the adiabatic Berry's connection ${\cal A}^{(n)}_{k}(\overline{P})$ associated with the specific $n$-th level of the fast system. 
 
 We made two basic approximations in deriving \eqref{effective Lagrangian-n}; the first is that we made the adiabatic approximation in \eqref{Auxialy expression} and the second is that the slower dynamical system was combined with the $n$-th level of the fast system by choosing $\epsilon_{n}$ suitably. This second operation does not exist in Berry's phase in general and specific to the anomalous Hall effect. Within these approximations, the Lagrangian \eqref{effective Lagrangian-n} may be said to be rather unique and obtained by the various ways of derivations, in addition to the original band theoretical derivation \cite{Niu}. The deformed commutation relations \eqref{modified commutators} would thus hold for \eqref{effective Lagrangian-n}, if one should attempt to define quantum mechanics based on \eqref{effective Lagrangian-n}. We take this deformation as an indication of the inconsistency of quantum mechanics to explain the anomalous Hall effect in the present setting. 
 Our model is an effective theory (and in that sense semi-classical) but our derivation of \eqref{effective Lagrangian-n} is based on the rules of quantum mechanics \cite{Berry}. 
 
 More generally, it is confirmed that the free Hamiltonian 
$H=\frac{P_{k}^{2}}{2M}$
is not re-written in terms of covariant derivatives $P_{k}\rightarrow P_{k}+eA_{k}(X)$ and $X_{k}\rightarrow X_{k}+{\cal A}_{k}(P)$ in Berry's phase approximation to give rise to \eqref{action} or 
\eqref{effective Lagrangian-n}; the covariant derivative of Berry's connection is given by $X_{k}+{\cal A}_{k}(P)$ with the gauge symmetry \eqref{gauge transformation2} in contrast to the electromagnetic covariant derivative $P_{k}+eA_{k}(X)$ in \eqref{covariant derivative1} \footnote{It is well known that $P_{k}+eA_{k}(X)$ is a covariant derivative in the X-representation where $X_{k}$ is an ordinary number and $P_{k}$ is a derivative. Similarly, $X_{k}+{\cal A}_{k}(P)$ is a covariant derivative in the P-representation where $P_{k}$ is an ordinary number and $X_{k}$ is a derivative.}. Generally the canonical Hamiltonian is defined in terms of covariant derivatives, namely, without using derivatives such as $\dot{X}_{k}$ or $\dot{P}_{k}$. Starting with the electromagnetic interactions
\begin{eqnarray}
H=\frac{(P_{k}+eA_{k}(X))^{2}}{2M} -e\phi(X),
\end{eqnarray}
one may define the canonical covariantization with respect to the gauge symmetry of Berry's connection by $X_{k}\rightarrow X_{k}+{\cal A}_{k}(P)$,
\begin{eqnarray}\label{Hamiltonian2}
H=\frac{(P_{k}+eA_{k}(X_{k}+{\cal A}_{k}(P)))^{2}}{2M} -e\phi(X_{k}+{\cal A}_{k}(P)).
\end{eqnarray}
But this is not the result \eqref{effective Lagrangian-n}. 
The action with the Hamiltonian \eqref{Hamiltonian2} differs from \eqref{effective Lagrangian-n}, as is seen by examining 
\begin{eqnarray}\label{Second order covariantization}
&&\int dt [P_{k}\dot{X}_{k}-H]\nonumber\\
&&=\int dt [P_{k}\dot{\overline{X}}_{k}+{\cal A}_{k}(P)\dot{P}_{k}
-\frac{(P_{k}+eA_{k}(\overline{X}))^{2}}{2M} +e\phi(\overline{X})]\nonumber\\
&&=\int dt [\overline{P}_{k}\dot{\overline{X}}_{k}-eA_{k}(\overline{X})\dot{\overline{X}}_{k}+{\cal A}_{k}(\overline{P}_{k}-eA_{k}(\overline{X}))(\dot{\overline{P}}_{k}-e\dot{A}_{k}(\overline{X}))\nonumber\\
&&\hspace{2cm} -\frac{\overline{P}_{k}^{2}}{2M} +e\phi(\overline{X})]
\end{eqnarray}
 where we set $\overline{X}_{k}=X_{k}+{\cal A}_{k}(P)$ in the second line and replaced $\overline{P}_{k}=P_{k}+eA_{k}(X)$ in the last line. The starting action $P_{k}\dot{X}_{k}-H$ in \eqref{Second order covariantization} defines the canonical commutation relations, but we do not obtain the action with Berry's phase \eqref{effective Lagrangian-n} or \eqref{action} up to a constant $\epsilon_{n}(p)= E_{n}(\overline{P})$ by a gauge covariantization to the order we show in \eqref{Hamiltonian2} \cite{Fujikawa2}. 
 
 In the formal level of covariantization, the variables $P_{k}$ in the Hamiltonian \eqref{Hamiltonian2} still need to be replaced by 
$P_{k}+eA_{k}(X)$ further; this procedure need to be continued {\em ad infinitum} to maintain the canonical commutation relations of the gauge invariant Hamiltonian.
One may conclude that 
the incompatibility of the gauge invariance of Berry's phase and the gauge invariance of the electromagnetic vector potential in the framework of quantum mechanics is thus regarded as a no-go theorem; one inevitably encounters the deviation as in \eqref{modified commutators} from the conventional canonical Hamiltonian formalism if one attempts to impose these two gauge symmetries simultaneously. The non-Hamiltonian character of \eqref{action} and \eqref{effective equations} is well known \cite{Niu2,Duval}.

The above consideration implies that the modification of the electron density of states by the applied magnetic field based on \eqref{action} and \eqref{modified commutators} with Berry's phase \cite{Niu2, Duval}, namely, the modification of the phase space volume $dV\rightarrow dV(1+ e\vec{B}\cdot \vec{\Omega})$ with $\vec{\Omega}$ standing for the Berry's curvature \cite{Sinitsyn}, is not defined in quantum mechanics, as naturally expected for a canonical system. 
 The quantum mechanical formalism in the presence of the electromagnetic vector potential is bound to end up with the non-quantum formalism, when one makes an adiabatic Berry's phase approximation.

To save the quantum mechanical description of the anomalous Hall effect, one may consider the action \eqref{effective Lagrangian-n} with the vanishing electromagnetic vector potential $A_{k}(X)=0$,
\begin{eqnarray}\label{actions}
S&=& \int dt \{\overline{P}_{k}\dot{X}_{k} +{\cal A}^{(n)}_{k}(\overline{P})\dot{\overline{P}_{k}}
-\epsilon_{n}(\overline{P}) + e\phi(X)\}\nonumber\\
&=& \int dt \{\overline{P}_{k}\dot{\overline{X}}_{k} 
- \epsilon_{n}(\overline{P}) + e\phi(\overline{X}_{k}+ {\cal A}^{(n)}_{k}(\overline{P}))\},
\end{eqnarray}
the second expression of which is consistent with the principle of quantum mechanics (i.e., canonical commutation relations and the covariant derivative) if one defines the covariant derivative $X_{k}=\overline{X}_{k}+{\cal A}^{(n)}_{k}(\overline{P})$, and the commutation relations
\begin{eqnarray}\label{canonical commutators}
[\overline{P}_{k},\overline{P}_{l}]=0, \
[\overline{P}_{k},\overline{X}_{l}]=-i\hbar\delta_{kl},\
[\overline{X}_{k},\overline{X}_{l}]=0.
\end{eqnarray}
 The presence of $e\phi(X)$ (we use $\phi(X)$ without any indices as the electromagnetic scalar potential) in the absence of the vector potential $A_{k}(X)$ is not very natural, but we ignore the unnaturalness by remembering a similar situation of the hydrogen atom. Physically, we set external $\vec{B}=0$ with $\vec{E}\neq0$. The possible operator ordering issue may arise here between $\overline{X}_{k}$ and $ {\cal A}^{(n)}_{k}(\overline{P})$, but we ignore it by noting that $ {\cal A}^{(n)}_{k}(\overline{P})$ is of the order of $O(\hbar)$ and thus to the accuracy of $O(\hbar)$ we are interested in the adiabatic approximation, it may be ignored \cite{Fujikawa2}. In general, the addition of an adiabatic Berry's connection combined with $\dot{\overline{P}_{k}}$ spoils the canonical structure of the slower system, but after a suitable operation, the term with $\dot{\overline{P}_{k}}$ is absorbed into the covariant derivative $\overline{X}_{k}+ {\cal A}^{(n)}_{k}(\overline{P})$ for $A_{k}(X)=0$, and the canonical structure is recovered as above. 
 
 The Hamiltonian in the canonical formalism in the second line of \eqref{actions} is given by 
\begin{eqnarray}\label{simplified Hamiltonian}
H=\epsilon_{n}(\overline{P})- e\phi(\overline{X}_{k}+ {\cal A}^{(n)}_{k}(\overline{P}))
\end{eqnarray}
 with the canonical commutation relations \eqref{canonical commutators}. The Hamiltonian is represented in terms of the covariant derivative $\overline{X}_{k}+ {\cal A}^{(n)}_{k}(\overline{P})$.
The Heisenberg equations of motion are 
\begin{eqnarray}\label{Hamilton equation3}
\dot{\overline{X}}_{k}=- \frac{\partial{\cal A}^{(n)}_{l}(\overline{P})}{\partial \overline{P}_{k}}\dot{\overline{P}} +\frac{\partial \epsilon_{n}(\overline{P})}{\partial \overline{P}_{k}}, \ \
\dot{\overline{P}}_{k}= e\frac{\partial}{\partial \overline{X}_{k}}\phi(\overline{X}_{l}+ {\cal A}^{(n)}_{l}(\overline{P})).
\end{eqnarray}
If one adopts the covariant derivative $X_{k}=\overline{X}_{k}+ {\cal A}^{(n)}_{k}(\overline{P})$, the equations of motion are given by \cite{Blount}
\begin{eqnarray}\label{Modified Berry's Phase}
\dot{X}_{k}=-\Omega^{(n)}_{kl}(\overline{P})\dot{\overline{P}}_{l} +\frac{\partial \epsilon_{n}(\overline{P})}{\partial \overline{P}_{k}}, \ \
\dot{\overline{P}}_{k}= e\frac{\partial}{\partial X_{k}}\phi(X),
\end{eqnarray}
where $\Omega^{(n)}_{kl}(\overline{P})$ stands for Berry's curvature defined by $ {\cal A}^{(n)}_{k}(\overline{P})$.
These are the common forms of the anomalous Hall effect \cite{Nagaosa1, Sinitsyn, Nagaosa}.
The commutation relations are also modified; using $X_{k}=\overline{X}_{k}+{\cal A}^{(n)}_{k}(\overline{P})$, the modified commutation relations are written as \cite{Blount}
 \begin{eqnarray}\label{modified canonical commutator}
[\overline{P}_{k},\overline{P}_{l}]=0, \
[\overline{P}_{k},X_{l}]=-i\hbar\delta_{kl},\
[X_{k},X_{l}]=i\hbar\Omega^{(n)}_{kl} .
\end{eqnarray}
The commutation relations written in the form \eqref{canonical commutators} are 
 canonical commutation relations. We can thus describe the anomalous Hall effect in quantum mechanics if one sets $A_{k}(x)=0$. But at the same time, the modification of the phase space volume, $dV(1+e\vec{B}\cdot \vec{\Omega})$ with $\vec{\Omega}$ standing for Berry's curvature, which may be used to describe the anomalous Nernst effect \cite{Sinitsyn}, is lost for $A_{k}(x)=0$.

The change of the Heisenberg equations from \eqref{Hamilton equation3} to the ``gauge invariant'' form \eqref{Modified Berry's Phase} is analogous to the change of equations in the purely electromagnetic Hamiltonian $H=(p_{l}+eA_{l}(x))^{2}/2M -e\phi(x)$ when one uses the covariant derivative $P_{l}=p_{l}+eA_{l}(x)$ in the Heisenberg equations. However, there is a difference in the present case of Berry's phase; the change of variables 
\begin{eqnarray}\label{variables change1}
X_{k}=\overline{X}_{k}+{\cal A}^{(n)}_{k}(\overline{P})
\end{eqnarray}
 changes the velocity of the electron, namely, the direction of the flow of the charge. If $\dot{X}_{k}$ in \eqref{Modified Berry's Phase} is transverse to the external force field up to $\frac{\partial \epsilon_{n}(\overline{P})}{\partial \overline{P}_{k}}$ since $\Omega^{(n)}_{kl}(\overline{P})\dot{\overline{P}}_{k}\dot{\overline{P}}_{l}=0$, $\dot{\overline{X}}_{k}$ in \eqref{Hamilton equation3} is not transverse. Usually, the gauge invariant forms of equations specify uniquely the current appearing there. 
However, the direction of the induced anomalous Hall current is not determined uniquely in the present approximate treatment, since the gauge symmetry of Berry's connection has apparently no relation with that of the electromagnetic vector potential.
If one should include the electromagnetic vector potential $A_{k}(X)$ the conserved electric current $e\dot{X}_{k}$ would be specified as the source current of electromagnetism, but then the quantum mechanical treatment would be spoiled by \eqref{modified commutators}.
We later argue that the Bjorken-Johnson-Low prescription may provide a solution to this technical issue by allowing essentially the action (1) with electromagnetic vector potential but with very small $F_{kl}$.
 
 In the analyses of the {\em spin Hall effect} \cite{Murakami,Sinova}, Berry's phase played an important role in \cite{Murakami}. The direct estimates of the spin Hall effect are based on the partial wave analysis induced by the spin-orbit coupling term but we prefer the estimates using Berry's phase in \cite{Murakami}. To be precise, the interplay of Berry's phase and the spin-polarization of the electron induced by the spin-orbit interaction is crucial in understanding the spin Hall effect. 
Applied to the spin Hall effect with a proper account of the spin-orbit coupling without referring to the electromagnetic field, the existing formulation \cite{Murakami} is consistent with the principle of quantum mechanics. It would be nice if one would give a simple way to show that the direction of the flow of the spin current is precisely transverse to the external force field $\dot{P}_{k}$ in quantum mechanical formalism.
The Bjorken-Johnson-Low prescription discussed below may give such an explanation by allowing the electromagnetic vector potential.

\section{Born-Oppenheimer approximation}
We examine the well-known formulation of adiabatic approximation, namely, the Born-Oppenheimer approximation \footnote{The Born-Oppenheimer approximation is not explicitly mentioned in \cite{Blount}, but some aspects discussed there without using the notion of Berry's phase are close to the Born-Oppenheimer approximation. For example, one may count the appearance of the covariant derivative $i\frac{d}{d\vec{P}} +{\cal A}_{k}(P)$ in \cite{Blount} (written in our notation).} and see if such an approximation would give a new insight into the description of the anomalous Hall effect. We start with the Hamiltonian with $eA_{k}=0$,
\begin{eqnarray}\label{Born-Oppenheimer Hamiltonian}
H=H_{0}(X_{k},P_{k}) + H_{1}(x_{k},p_{k};P_{k}).
\end{eqnarray}
 The fast system is described by $H_{1}(x_{k},p_{k};P_{k})$ as before and the slower system is described by $H_{0}(X_{k},P_{k})$ but with $H_{0}(X_{k},P_{k})=\frac{P_{k}^{2}}{2M} +V(X_{k})$. The difference from \eqref{starting Hamiltonian2} is that we do not include the electromagnetic scalar potential but instead we include a general scalar potential $V(X_{k})$; we set $V(X_{k})=-e\phi(X_{k})$ at the end, where $\phi(X_{k})$ without any index stands for the electromagnetic scalar potential. This is a technical procedure to avoid the issue of electromagnetic gauge symmetry as much as possible. We set the electromagnetic vector potential $eA_{k}=0$ from the beginning since it spoils the Hamiltonian formalism in \eqref{effective Lagrangian-n} (and in a related formula in the present formulation) when one introduces the quantity corresponding to Berry's connection in the context of adiabatic formulation; see \eqref{Second order covariantization}. 
 
 The total wave function of the Born-Oppenheimer approximation is assumed to be of the form \cite{Born} 
\begin{eqnarray}\label{total wave function}
\Psi(x_{k},P_{k})=\sum_{l}\varphi_{l}(P_{k})\phi_{l}(x_{k},P_{k})
\end{eqnarray}
where $\phi_{l}(x_{k},P_{k})$ is defined in the present approximation
\begin{eqnarray}\label{instantaneous solution}
H_{1}(x_{k},p_{k};P_{k})\phi_{l}(x_{k},P_{k})=E_{l}(P)\phi_{l}(x_{k},P_{k}),
\end{eqnarray}
namely, it is chosen as an instantaneous eigenfunction of the fast system $H_{1}(x_{k}, p_{k}; P_{k})$ as in \eqref{instantaneous function} but without the electromagnetic vector potential. The Born-Oppenheimer approximation is based on quantum mechanics and thus on the canonical commutation relations
 \begin{eqnarray}\label{canonical commutators2}
 [P_{k},P_{l}]=0, \ [P_{k},X_{l}]=-i\delta_{kl},\ [X_{k},X_{l}]=0.
\end{eqnarray}
We choose the representation of $P_{k}$ diagonal (P-representation) for the slower system. 
The Born-Oppenheimer approximation \eqref{total wave function} is an infinite number of coupled equations unlike the Berry's phase approximation. A way to cope with this difference may be to pick up specific states in both slower and faster systems, namely, to pick up one of the combinations in \eqref{total wave function}
\begin{eqnarray}\label{Single state formula}
\Psi_{l}=\varphi_{l}(P)\phi_{l}(x;P)
\end{eqnarray}
where $l$ is left arbitrary in the present treatment; one may set $l=1$, for example. 
The use of the full version \eqref{total wave function} shall be discussed later; the adiabatic approximation becomes more visible in the full version.
When the wave function \eqref{Single state formula} is operated to the Hamiltonian \eqref{Born-Oppenheimer Hamiltonian} in the Schr\"{o}dinger equation $(H_{0}+H_{1})\Psi_{l}(x_{k},P_{k})=E\Psi_{l}(x_{k},P_{k})$ and after multiplying the both sides by $\phi^{\ast}_{l}(x_{k},P_{k})$ and integrating over $\int d^{3}x$, one obtains 
\begin{eqnarray}\label{Hamiltonian formalism}
&&{\frac{P_{k}^{2}}{2M}\varphi_{l}(P_{k})+\int d^{3}x\phi_{l}^{\ast}(x_{k},P_{k})V(X_{k})\phi_{l}(x_{k},P_{k})}\varphi_{l}(P_{k})+E_{l}(P_{k})\varphi_{l}(P_{k})\nonumber\\
&&\hspace{7cm}=E\varphi_{l}(P_{k})
\end{eqnarray}
where we assumed the instantaneous solutions \eqref{instantaneous solution} with the assumed orthonormality
$\int d^{3}x\phi^{\ast}_{l}(x;P)\phi_{n}(x;P)=\delta_{nl}$.
Using the completeness relation $\sum_{l^{\prime}}\phi_{l^{\prime}}(x,P)\phi_{l^{\prime}}^{\ast}(y,P)=\delta(x-y)$, one has 
\begin{eqnarray}\label{matrix notation}
&&\int d^{3}x\phi_{l}^{\ast}(x,P)V(X)\phi_{l}(x,P)=V_{0}+
V_{1}^{k}\int d^{3}x\phi_{l}^{\ast}(x,P)X_{k}\phi_{l}(x,P)\nonumber\\
&&+ V_{2}\sum_{l^{\prime}}\int d^{3}xd^{3}y\phi_{l}^{\ast}(x,P)X_{k}\phi_{l^{\prime}}(x,P)\phi^{\ast}_{l^{\prime}}(y,P)X_{k}\phi_{l}(y,P)+ \cdots \nonumber\\
&=&V_{0}+V^{k}_{1}[X_{k}+{\cal A}^{ll}_{k}] +V_{2}\sum_{l^{\prime}}[\delta_{ll^{\prime}}X_{k}+{\cal A}^{ll^{\prime}}_{k}][\delta_{l^{\prime}l}X_{k}+{\cal A}^{l^{\prime}l}_{k}]+\cdots
\end{eqnarray}
if one assumes a power expansion of $V(X)=V_{0}+V^{k}_{1}X_{k}+ V_{2}X_{k}X_{k}+ ...$ . We defined 
\begin{eqnarray}\label{Berry connection2}
{\cal A}^{ln}_{k}(P_{m}) =\int d^{3}x\phi_{l}^{\ast}(x_{m},P_{m})X_{k}\phi_{n}(x_{m},P_{m})
\end{eqnarray}
and $X_{k}$ stands for the linear operator 
$X_{k}=i\frac{\partial}{\partial P^{k}}$ which acts on $\phi_{n}(x_{m},P_{m})$ in \eqref{Berry connection2}; it also acts on $\varphi_{l}(P)$ in \eqref{Hamiltonian formalism} and also on ${\cal A}^{l^{\prime}l}_{k}(P)$ appearing on the right hand side of it in \eqref{matrix notation}. The relation \eqref{matrix notation} shows that one obtains the mixing of other fast states if $V(X)$ contains second and higher powers in $X_{k}$. To avoid the mixing, we here assume a linear form of $V(X_{k})$ in $X_{k}$, which implies that the external force field is linear in the electric field $\vec{E}= {\rm constant}$ after we replace $V(X)\rightarrow -e\phi(X)$; $eV^{k}_{1}$ corresponds to the constant external electric field and this approximation may still be realistic. In the above linear field approximation, we have 
\begin{eqnarray}\label{Diagonal matrix element} 
\int d^{3}x\phi_{l}^{\ast}(x,P)V(X)\phi_{l}(x,P) = V(X_{k}+ {\cal A}_{k}^{(l)}(P))
\end{eqnarray}
where ${\cal A}_{k}^{(l)}(P)\equiv{\cal A}^{ll}_{k}(P)$.

Using \eqref{Diagonal matrix element} in \eqref{Hamiltonian formalism} one obtains 
\begin{eqnarray}\label{diagonal dominance}
\int d^{3}x\phi_{l}^{\ast}(x,P)V(X_{k})\phi_{l}(x,P)\varphi_{l}(P)
&=&V(X_{k}+ {\cal A}^{(l)}_{k}(P))\varphi_{l}(P).
\end{eqnarray}
 To be precise, we formulate for a general scalar potential $V(X_{k})$ first and identify later the scalar potential with the electromagnetic scalar potential $V(X_{k})=-e\phi(X_{k})$. We thus have the Schr\"{o}dinger equation
 \begin{eqnarray}\label{Hamiltonian equation}
 \{\epsilon_{l}(P_{m})-e\phi(X_{k}+ {\cal A}^{(l)}_{k}(P_{m}))\}\varphi_{l}(P)=E\varphi_{l}(P)
 \end{eqnarray}
with $\epsilon_{l}(P_{m})=\frac{P_{m}^{2}}{2M}+E_{l}(P_{m})$, namely, the Born-Oppenheimer effective Hamiltonian for the slower system is given by 
\begin{eqnarray}\label{Hamiltonian3}
H_{l}(P)= \epsilon_{l}(P)-e\phi\left(X_{k}+{\cal A}^{(l)}_{k}(P) \right) 
\end{eqnarray}
which acts on the wave function $\varphi_{l}(P)$. We here made the linear field assumption, namely, the external force field is $\vec{E}={\rm constant}$ with $e\phi(X_{k})$ linear in $X_{k}$.
This final formula of the Hamiltonian \eqref{Hamiltonian3} agrees with Berry's phase formalism \eqref{actions} including canonical commutation relations \eqref{canonical commutators2} and the gauge symmetry associated with Berry's connection, although the present formula is valid only in the linear field approximation of $e\phi(X_{k})$.

In the Born-Oppenheimer approximation, we start with an infinite dimensional expression \eqref{total wave function}. But to relate the final result to Berry's phase formulation, we tentatively assumed that a specific slow system $\varphi_{l}(P)$ mainly couples to a specific fast system as in $\Psi_{l}=\varphi_{l}(P)\phi_{l}(x;P)$,
and we obtain the Hamiltonian \eqref{Hamiltonian3} which is the same as Berry's phase formulation of the anomalous Hall effect, in the linear field approximation and in the absence of the electromagnetic vector potential. 
It is interesting that we get the identical formula for Berry's connection in either formulation; note that Berry's connection (with $A_{k}(X)=0$) is determined uniquely by \eqref{instantaneous solution}.

We now examine the original Born-Oppenheimer approximation defined with an infinite number of states \eqref{total wave function}.
In the formulation of Berry's phase in Section 2, one obtained Berry's phase by a diagonal approximation of the slowly varying quantity ${\cal A}^{(ln)}_{k}(P)\dot{P}_{k}$ based on the well-defined adiabatic formula and, in principle, for any form of the electromagnetic scalar potential $-e\phi(X)$. In the Born-Oppenheimer approximation, if one starts with an infinite number of states \eqref{total wave function}, one obtains extra terms 
\begin{eqnarray}\label{extra term}
V_{1}^{k} \sum_{n\neq l}{\cal A}^{(ln)}_{k}(P)\varphi_{n}(P)
\end{eqnarray}
on the right hand side of \eqref{diagonal dominance} in the linear field approximation; to be precise, one has $\sum_{n}\phi_{n}(x,P)\varphi_{n}(P)$ instead of $\phi_{l}(x,P)\varphi_{l}(P)$ on the left hand side of \eqref{diagonal dominance} and the extra \eqref{extra term} on the right hand side of \eqref{diagonal dominance}.

In an explicit form, the Schr\"{o}dinger equation \eqref{Hamiltonian equation} in the presence of \eqref{extra term} is re-written as an infinite dimensional matrix relation 
\begin{eqnarray}\label{matrix notation2}
{\cal M}(V_{1}^{k}{\cal A}^{(ln)}_{k}(P), H_{l})\psi(P)=E\psi(P)
\end{eqnarray}
with 
\begin{eqnarray}\label{Matrix}
{\cal M}(V_{1}^{k}{\cal A}^{(ln)}_{k}(P), H_{l})=\left(\begin{array}{ccccc}
H_{1}& V_{1}^{k}{\cal A}^{(12)}_{k}(P)&V_{1}^{k}{\cal A}^{(13)}_{k}(P)&V_{1}^{k}{\cal A}^{(14)}_{k}(P)&...\\
V_{1}^{k}{\cal A}^{(21)}_{k}(P)&H_{2}&V_{1}^{k}{\cal A}^{(23)}_{k}(P)&V_{1}^{k}{\cal A}^{(24)}_{k}(P)&...\\
V_{1}^{k}{\cal A}^{(31)}_{k}(P)&V_{1}^{k}{\cal A}^{(32)}_{k}(P)&H_{3}&V_{1}^{k}{\cal A}^{(34)}_{k}(P)&...\\
...&...&...&...&...\\
\end{array}\right)
\end{eqnarray}
and an infinite dimensional vector
\begin{eqnarray}\label{infinite dimensional vector}
\psi^{T}(P)= (\varphi_{1}(P), \varphi_{2}(P), \varphi_{3}(P), \varphi_{4}(P),... )
\end{eqnarray}
where $T$ stands for the transpose of a column vector, and $H_{l}$ is defined in \eqref{Hamiltonian3}. One still assumes the linear field approximation of $e\phi(X)$.
One may then recognize that the relation \eqref{Hamiltonian equation} corresponds to a diagonal dominance (adiabatic) approximation with very small off- diagonal elements 
$V_{1}^{k}{\cal A}^{(ln)}_{k}(P_{m})$, $l\neq n$, set to zero
in the relation \eqref{matrix notation2}; the diagonal dominance approximation is valid up to the accuracy of 
the square of $V_{1}^{k}{\cal A}^{(ln)}_{k}(P_{m})$, as is confirmed by considering the determinant of \eqref{Matrix}, for example.

In Berry's phase approximation in Section 2, one had slow and very small quantity ${\cal A}^{(ln)}_{k}(P)\dot{P}_{k}$ which justified the adiabatic approximation \eqref{simplified Hamiltonian}; the diagonal dominance approximation was applied to the faster system. In contrast, one needs to have slow and very small $V_{1}^{k}{\cal A}^{(ln)}_{k}(P_{m})$ in the linear field approximation of the potential in the present Born-Oppenheimer approximation \eqref{matrix notation2}; the diagonal dominance approximation was applied to the slower system. One may say that the case of Berry's phase and its adiabatic approximation are better defined mathematically, although one still had to set $A_{k}(X)=0$ in \eqref{actions} to maintain the consistency with quantum mechanics.

\section{Bjorken-Johnson-Low prescription} 
Following the conventional understanding, we have considered in the main part of this paper so far that the commutation relations are defined by the Lagrangian with the adiabatic approximation instead of the starting original Lagrangian, except for \eqref{canonical commutators2}. The adiabatic approximation in general may define the commutation relations different from the canonical ones. We then examined how the canonical commutation relations are realized for the (simplified) adiabatic Lagrangian; to be precise, we asked if the canonical commutation relations are preserved for the slower system with the vanishing electromagnetic vector potential.

 On the other hand, following the Bjorken-Johnson-Low (BJL) prescription \cite{BJL} and assuming that one started with a well-defined quantum mechanical system as in our case, one may understand that the commutation relations of both slower and fast systems are always defined by the original Lagrangian which describes precisely the high frequency behavior of the original theory \cite{Deguchi2}. The BJL prescription defines the commutation relation such as $[P_{k}(0),X_{l}(0)]$ based on the examination of the high frequency behavior (i.e., $\omega\rightarrow \infty$ ) of the correlation function ($T^{\ast}$ standing for the time-ordering operation) 
 \begin{eqnarray}
 \int dt e^{i\omega t}\langle T^{\ast}P_{k}(t)X_{l}(0)\rangle
 \end{eqnarray}
 in path integrals, for example, and the time derivative terms in the action determine the commutation relations; it explains why the commutation relations in $H=P_{k}^{2}/2M +V(X)$ when written in terms of the associated Lagrangian $L=P_{k}\dot{X}_{k}-H$ are independent of the size of the finite potential $V(X)$. The BJL prescription is important to define (inevitably modified) commutation relations in the presence of quantum anomalies, for example. A brief account of the BJL prescription is given in \cite{Fujikawa2}. 
 
 We first define the quantum mechanical system by \eqref{starting Hamiltonian2} with canonical commutation relations \eqref{standard commutator-0} and \eqref{standard commutator-2} and then analyze the adiabatic approximations which produce Berry's phase. The important fact is that the high frequency behavior of the original Lagrangian can be very different from that of the adiabatic Lagrangian with Berry's phase. The adiabatic approximation describes the theory at the slow moving limit of the parameter in \eqref{Auxialy expression}, in the present case $\dot{P}_{k}$, and thus throwing away many terms with $\dot{P}_{k}$, while the high frequency limit of $\dot{P}_{k}$ describes the original theory by retaining all the terms with $\dot{P}_{k}$. As a result, the original theory at the high frequency limit of $\dot{P}_{k}$ becomes independent of $\dot{P}_{k}$ after a summation over all the terms with
 $\dot{P}_{k}$ and giving rise to the starting full theory \eqref{starting Hamiltonian2}.
The high frequency behavior of the adiabatic Lagrangian may be said to be superficial and not generally realized by a smooth deformation
starting with the original Lagrangian that is defined by canonical commutation relations. For example, the first expression of \eqref{Auxialy expression} gives the precise high frequency formula equivalent to the starting Lagrangian \eqref{starting Hamiltonian2} after summing over $n$ and $l$, which is thus defined by the canonical commutation relations, while the second expression of \eqref{Auxialy expression} gives superficial deformed commutation relations. 
 
 The effect described by the adiabatic approximation such as the anomalous Hall effect 
 is perfectly physical in quantum mechanics but the associated deformed commutation relations, which are determined by the high frequency limits of the adiabatic Lagrangian, may not be reliable. As an analogous effect, one may remember that the low-energy rotation property of the electron spin is modified by Schwinger's anomalous magnetic moment in QED (and in fact this is the way how the anomalous magnetic moment is measured), but the equal-time commutation relations of the electron field, which are determined by the high frequency behavior, are not modified by the induced anomalous magnetic moment. The electron commutation relations are determined precisely by the starting QED.
 
 If one should adopt the BJL prescription, none of the effects associated with the deformed commutation relations would appear in principle. The inevitable appearance of the canonical commutation relations in the BJL prescription would make the analysis of \eqref{modified commutators} unnecessary. 
 At the same time, the simultaneous appearance of Berry's phase and the electromagnetic vector potential, that has been excluded in the main part of the present paper since these two put together are judged to be incompatible with the canonical commutation relations in the {\em adiabatic} Hamiltonian as shown in \eqref{Second order covariantization}, would now be allowed. One does not ask the commutation properties of \eqref{action} and \eqref{effective equations} in the adiabatic limit, since the commutation relations are not defined in the adiabatic domain in the sense of BJL prescription which defines the commutation relations in the high frequency limits. As for the check of the quantum mechanical consistency, one uses the starting equations \eqref{starting Hamiltonian2} instead of \eqref{Second order covariantization}; one would then recognize that the starting property of \eqref{starting Hamiltonian2} is deformed by the adiabatic formula as a result of approximation. The important point is that Berry's phase does not modify the basic quantum commutation properties \eqref{standard commutator-0} and \eqref{standard commutator-2} \cite{Berry}. 
 The difference between the adiabatic form and the non-adiabatic form of the Lagrangian in the presence of Berry's phase shall be illustrated for a specific model of the two-level crossing in the Appendix.

 \section{Conclusion}
Originally, Berry's phase is defined as a response of a (fast) quantum mechanical system to an infinitely slow variation of the background variables \cite{Berry}. To regard those slow background variables as actually a part of the dynamical variables of a slower dynamical system, which is coupled to a fast system, is a new and ingenious feature of the theory of the anomalous Hall effect. But we encounter the problem such as the possible modification of commutation relations of the slower variables by Berry's phase. We have examined to what extend we can analyze the anomalous Hall effect as a quantum mechanical problem. 
 
 In the case of Berry's phase we analyzed, we have shown that the Hamiltonian of the slower system without the electromagnetic vector potential is summarized in the form of \eqref{simplified Hamiltonian} 
 (we actually suppressed the excessive appearance of characters with overbars in \eqref{simplified Hamiltonian})
\begin{eqnarray}
H=\epsilon_{n}(P)- e\phi(X_{k}+ {\cal A}^{(n)}_{k}(P))
\end{eqnarray}
and the canonical commutation relations
\begin{eqnarray}
[P_{k},P_{l}]=0, \
[P_{k},X_{l}]=-i\hbar\delta_{kl},\
[X_{k},X_{l}]=0.
\end{eqnarray}
with the Heisenberg equations of motion 
\begin{eqnarray}\label{equations in conclusion1}
\dot{X}_{k}=-\frac{\partial {\cal A}^{(n)}_{m}(\vec{P})}{\partial P_{k}}\dot{P}_{m} +\frac{\partial \epsilon_{n}(\vec{P})}{\partial P_{k}}, \ \ \
\dot{P}_{k}= e\frac{\partial}{\partial X_{k}}\phi(\vec{X}+ \vec{{\cal A}}^{(n)}(\vec{P})).
\end{eqnarray}
If one uses the covariant derivatives
 $\overline{X}_{k}=X_{k}+{\cal A}^{(n)}_{k}(P)$ \cite{Blount}, the conventional ``gauge invariant'' (with respect to Berry's phase) equations of motion are obtained 
\begin{eqnarray}\label{equations in conclusion}
\dot{\overline{X}}_{k}=-\Omega^{(n)}_{km}(\vec{P})\dot{P}_{m} +\frac{\partial \epsilon_{n}(\vec{P})}{\partial P_{k}}, \ \ \
\dot{P}_{k}= e\frac{\partial}{\partial \overline{X}_{k}}\phi(\vec{\overline{X}}),
\end{eqnarray}
with the modified commutation relations 
 \begin{eqnarray}
[P_{k},P_{l}]=0, \
[P_{k},\overline{X}_{l}]=-i\hbar\delta_{kl},\
[\overline{X}_{k},\overline{X}_{l}]=i\hbar\Omega^{(n)}_{kl} .
\end{eqnarray}
The current $\dot{\overline{X}}_{k}$ in \eqref{equations in conclusion} is transverse to the applied external force $\dot{P}_{k}$
since $\Omega^{(n)}_{km}(\vec{P})\dot{P}_{m}\dot{P}_{k}=0$ up to the term $\frac{\partial \epsilon_{n}(\vec{P})}{\partial P_{k}}$. 
Eq.\eqref{equations in conclusion} is ``gauge invariant'' with respect to Berry's phase and thus preferred by the analysis of Berry's phase. But the gauge symmetries of the electromagnetism and Berry's phase have no connection, in principle, in the absence of electromagnetic vector potential; this absence is required since the electromagnetic vector potential 
would lead to the difficulty \eqref{modified commutators} in quantum mechanics. The change of variables in \eqref{equations in conclusion1} and \eqref{equations in conclusion}
 \begin{eqnarray}\label{change of variables2}
\dot{X}_{k}=\dot{\overline{X}}_{k}-\dot{{\cal A}}^{(n)}_{k}(\vec{P}) .
 \end{eqnarray}
indicates that the change of variables from gauge non-invariant $\dot{X}_{k}$ to invariant $\dot{\overline{X}}_{k}$ with respect to Berry's phase changes the direction of the induced electric current. Under the transformations \eqref{change of variables2} in the present notation, the electromagnetic properties basically stay unchanged. In the framework of Berry's phase, we failed to show directly this technical issue; namely, the induced anomalous Hall current is unambiguously transverse to the applied external force field (although it is not clear if such a precise transversality is physically required) in a hypothetical situation of the absence of the electromagnetic vector potential.

The main aspects of the anomalous Hall effect \eqref{equations in conclusion} are explained by Berry's phase but one may conclude that the equations \eqref{action}, \eqref{effective equations} and \eqref{modified commutators} including the description of the anomalous Nernst effect are not fully realized in the framework of quantum mechanics because of \eqref{modified commutators} and \eqref{Second order covariantization}.

On the other hand, if one should adopt the Bjorken-Johnson-Low (BJL) prescription, which differs from the idea of commutation relations defined by the adiabatic Lagrangian, one would obtain a different picture. Basically, an analogue of Schwinger's anomalous magnetic moment is realized, as already mentioned briefly.
 The relations of \eqref{action} and \eqref{effective equations}, which are the results of the adiabatic approximation to the original Lagrangian defined in terms of the Hamiltonians \eqref{starting Hamiltonian2}, would be defined in the adiabatic domain where the action principle for \eqref{action} is expected to be valid in the limited sense. The relations \eqref{action} and \eqref{effective equations} would form a valid set of solutions combined with the starting canonical commutation relations \eqref{standard commutator-0} and \eqref{standard commutator-2}. In this picture, the adiabatic Berry's phase (the monopole-like object) would disappear in the high frequency non-adiabatic domain.
 
If one should adopt this view of BJL, one would obtain a consistent picture of the existing Berry's phase theory of the anomalous Hall effect in the following manner: The electromagnetic field $A_{k}$ may now be included in the present adiabatic Lagrangian, and the coordinates $\dot{\overline{X}}_{k}$ in \eqref{equations in conclusion} correspond to the source current of $A_{k}$, namely, $\dot{x}_{k}$ in \eqref{action} or $\dot{X}_{k}$ in \eqref{effective Lagrangian-n}; the direction of the induced electric flow is then transverse to the external force field as in \eqref{effective equations} in the adiabatic domain. (Even with the presence of $A_{k}(x)$ in \eqref{effective equations}, however, the transverse velocity is orthogonal to $\dot{p}_{k}$, which is a composed force of two terms on the right hand side of the second equation of \eqref{effective equations}, and not the external force $ e\frac{\partial}{\partial x_{k}}\phi(\vec{x})$ itself. ) The modification of the phase space volume, which is derived from a careful analysis of the adiabatic equations \eqref{effective equations} as in \cite{Niu2}, would explain the anomalous Nernst effect as a low energy adiabatic effect \cite{Sinitsyn}; the low energy here means the energy domain where Berry's connection is well-defined. Then one would obtain a consistent picture of all the main features of adiabatic Berry's phase theory of the anomalous Hall effect. 

The possible explanation of quantum anomalies in terms of Berry's phase appearing in the anomalous Hall effect, for which the deformation of the canonical commutation relations by the external magnetic field $\vec{B}$ is essential, is not possible in the framework of quantum mechanics, either using the adiabatic Lagrangian for which no $\vec{B}$ is allowed or using the BJL prescription for which only the canonical commutation relations appear \cite{Fujikawa2}. The quantum anomalies, as short distance effects, are generally characterized by the following fact; the symmetries which contain anomalies lead to deformed commutation relations (symmetries) at high frequency limits when analyzed by the BJL prescription.
 
 \section*{Acknowledgements}One of us (KF) is supported in part by JSPS
KAKENHI (Grant No.18K03633), and he is grateful to N. Nagaosa for a discussion at the early stage of the present study. The other (KU) is supported in part by Grant for the Academic Prize of College of Science and Technology, Nihon University; he thanks S. Murakami for helpful discussions. 

\appendix 

\section{Two-level system}
In this appendix we illustrate the properties related to the BJL prescription \cite{BJL} using the two-level crossing Hamiltonian, which does not contain the electromagnetic vector potential. We study the so-called Weyl-type level crossing 
\begin{eqnarray}
H_{W}= \mu \vec{\sigma}\cdot \vec{P}(t)
\end{eqnarray}
 where $\mu$ is a suitable constant and $\vec{\sigma}$ are the Pauli matrices, which stand for the pseudo-spin not for the real spin, with the time dependent 
 \begin{eqnarray}\label{original form}
 &&\vec{P}(t)=|P|(\sin\theta\cos\varphi,\sin\theta\sin\varphi, \cos\theta ),\nonumber\\
 &&\vec{\sigma}\cdot\vec{P}=|P|\left(\begin{array}{cc}
 \cos\theta&e^{-i\varphi}\sin\theta\\
 e^{i\varphi}\sin\theta&-\cos\theta
 \end{array}\right).
 \end{eqnarray}
 The action of the second quantization is given by 
 \begin{eqnarray}\label{second quantization}
 S&=&\int dt [\hat{\psi}^{\ast}(t)i\hbar\partial_{t}\hat{\psi}(t)- \hat{\psi}^{\ast}(t)H_{W}\hat{\psi}(t)]\nonumber\\
 &=& \int dt [\sum_{k=\pm}\hat{a}_{k}^{\ast}(t)i\hbar\partial_{t} \hat{a}_{k}(t)
 + \sum_{k,l}\hat{a}_{k}^{\ast}(t)(u_{k}^{\ast}i\hbar\partial_{t} u_{l})\hat{a}_{l}(t)
 -\sum_{k=\pm}\hat{a}_{k}^{\ast}(t)E_{k}\hat{a}_{k}(t)]\nonumber\\
 &=&\int dt [\hat{a}^{\ast}(t)i\hbar\partial_{t} \hat{a}(t)
-\hat{a}^{\ast}(t)H^{\prime}\hat{a}(t)]
\end{eqnarray}
where we used the instantaneous (snapshot) approximation with the instantaneous eigenvalues $\pm \mu |P|(t)$
\begin{eqnarray}
&&\hat{\psi}(t)=\sum_{k=\pm}\hat{a}_{k}(t)u_{k}(t), \ \ \
\mu \vec{\sigma}\cdot \vec{P}(t)u_{\pm}(t)=E_{\pm}u_{\pm}(t)=\pm\mu |P|(t)u_{\pm}(t),\nonumber\\
&&u_{+}=\left(\begin{array}{c}
e^{-i\varphi}\cos\frac{1}{2}\theta \\
\sin\frac{1}{2}\theta
\end{array}\right), \ \ \
u_{-}=\left(\begin{array}{c}
e^{-i\varphi}\sin\frac{1}{2}\theta \\
-\cos\frac{1}{2}\theta
\end{array}\right)
\end{eqnarray}
and one obtains the conveniently parameterized Hamiltonian defined by $u_{\pm}(t)$
\begin{eqnarray}\label{parameterized Hamiltonian}
H^{\prime}&=& \left(\begin{array}{cc}
E_{+}-u_{+}^{\ast}i\hbar\partial_{t}u_{+}&-u_{+}^{\ast}i\hbar\partial_{t}u_{-} \\ -u_{-}^{\ast}i\hbar\partial_{t}u_{+}&E_{-}-u_{-}^{\ast}i\hbar\partial_{t}u_{-}\end{array}\right)\nonumber\\
&=&\left(\begin{array}{cc}
\mu |P|&0\\ 0&-\mu |P|\end{array}\right) - \hbar\left(\begin{array}{cc}
\frac{(1+\cos\theta)}{2}&\frac{\sin\theta}{2}\\
 \frac{\sin\theta}{2}&
\frac{(1-\cos\theta)}{2}
 \end{array}\right)\dot{\varphi}
-\hbar\left(\begin{array}{cc}
0&i\\
-i&0
 \end{array}\right)\frac{1}{2}\dot{\theta}
\end{eqnarray}
with 
\begin{eqnarray}
\hat{a}(t)=\left(\begin{array}{c}
\hat{a}_{+}(t) \\ \hat{a}_{-}(t)
\end{array}\right).
\end{eqnarray}
The exact snapshot solution of the Schr\"{o}dinger equation starting with the second quantization defined by $\psi_{l}(T)=\langle 0|\hat{\psi}(T) \hat{a}^{\dagger}_{l}(0)|0\rangle$ by solving $i\hbar\partial_{t}\hat{a}(t)=H^{\prime}\hat{a}(t)$ to be used for $\hat{\psi}(T) $, is then written in the form 
\begin{eqnarray}\label{state vector}
\psi_{l}(T)=\sum_{k=\pm}
 u_{k}(T)\langle k|T^{\ast}\exp\{ -\frac{i}{\hbar}\int_{0}^{T} dt\left(\begin{array}{cc}
E_{+}-u_{+}^{\ast}i\hbar\partial_{t}u_{+}&-u_{+}^{\ast}i\hbar\partial_{t}u_{-} \\ -u_{-}^{\ast}i\hbar\partial_{t}u_{+}&E_{-}-u_{-}^{\ast}i\hbar\partial_{t}u_{-}\end{array}\right)\}|l\rangle 
\end{eqnarray}
with the initial condition $\psi_{l}(0)=u_{l}(0)$ and $T^{\ast}$ stands for the time ordering operation; the vectors $|k\rangle=\hat{a}_{k}^{\dagger}(0)|0\rangle$ with $k=\pm$. 
If one assumes the adiabatic configurations
\begin{eqnarray}\label{constraint}
\mu|P|/\hbar\gg |\frac{\sin\theta}{2}\dot{\varphi}| \sim |\frac{(1\pm\cos\theta)}{2}\dot{\varphi}|, \ \ \ \mu|P|/\hbar\gg |\dot{\theta}|
\end{eqnarray}
the solution of the Schr\"{o}dinger equation by the adiabatic approximation is given for a periodic system by a diagonal form (see, for example, \cite{Nagaosa1})
\begin{eqnarray}\label{state vector2}
\psi_{l}(T)= \sum_{k=\pm}u_{k}(T)\langle k| T^{\ast}\exp\{ -i\int_{0}^{T} dt\left(\begin{array}{cc}
\frac{\mu |P|}{\hbar}-\frac{(1+\cos\theta)}{2}\dot{\varphi}&0 \\ 0&-\frac{\mu |P|}{\hbar}-\frac{(1-\cos\theta)}{2}\dot{\varphi}\end{array}\right)\}|l\rangle
\end{eqnarray}
The $\varphi$-dependent terms give the monopole-like topological terms (i.e., Berry's phase with $e_{M}=2\pi \hbar$, for the periodic boundary condition $\varphi(0)=0$ and $\varphi(T)=2\pi$)
\begin{eqnarray}
2\pi\hbar\int_{0}^{T}\frac{(1\pm\cos\theta)}{4\pi}\dot{\varphi} dt=e_{M}\int_{0}^{T}\frac{(1\pm\cos\theta)}{4\pi(|P|\sin\theta)}|P|\sin\theta\dot{\varphi}dt=e_{M}\oint {\cal A}^{\pm}_{\varphi}ds_{\varphi}
\end{eqnarray}
with the line element $ds_{\varphi}=|P|\sin\theta\dot{\varphi}dt$, or in a more common notation (in the main text, we normalized $e_{M} {\cal A}^{\pm}_{k}\rightarrow {\cal A}^{\pm}_{k}$)
\begin{eqnarray}
\int_{0}^{T}u_{\pm}^{\ast}i\hbar\partial_{t}u_{\pm}dt=e_{M}\int_{0}^{T} {\cal A}^{\pm}_{k}\dot{P}^{k}dt.
\end{eqnarray}
In the adiabatic domain \eqref{constraint}, the monopoles appear near the level-crossing point. 
The monopole singularities disappear at away from the adiabatic domains
$|\frac{(1\pm\cos\theta)}{2}\dot{\varphi}|\gg\mu|P|/\hbar$; 
this is seen by applying the transformation which diagonalizes the coefficient matrix of $\dot{\varphi}$, 
\begin{eqnarray}
\hat{a}=\left(\begin{array}{c}
\hat{a}_{+} \\ {a}_{-}
\end{array}\right)\rightarrow \hat{a} =
\left(\begin{array}{cc}
\cos\frac{1}{2}\theta & -\sin\frac{1}{2}\theta \\ 
\sin\frac{1}{2}\theta& \cos\frac{1}{2}\theta
\end{array}\right)
\left(\begin{array}{c}
\hat{a}^{\prime}_{+} \\ \hat{a}^{\prime}_{-}
\end{array}\right)
\end{eqnarray}
in \eqref{second quantization}. The coefficient matrix of $\dot{\varphi}$ in \eqref{parameterized Hamiltonian} has a unit determinant and a unit trace and thus the eigenvalues of $1\ {\rm and}\ 0$ independent of $\theta$. The state vector \eqref{state vector} and the action appearing in the shoulder are then replaced by the terms with the coefficient of $\dot{\varphi}$ diagonal
\begin{eqnarray}\label{modified state}
&&\psi^{\prime}_{l}(T)= \sum_{k}u^{\prime}_{k}(T)\nonumber\\
&&\times\langle k| T^{\ast}\exp\{ -\frac{i}{\hbar}\int_{0}^{T} dt[\mu |P|\left(\begin{array}{cc}
\cos\theta&-\sin\theta\\ -\sin\theta&-\cos\theta
\end{array}\right)-\left(\begin{array}{cc}
1&0\\ 0&0
\end{array}\right)\hbar\dot{\varphi}
]\}|l\rangle.
\end{eqnarray}
where $u^{\prime}_{+}(T)=u_{+}(T)\cos\frac{1}{2}\theta + u_{-}(T)\sin\frac{1}{2}\theta$ and $u^{\prime}_{-}= -u_{+}(T)\sin\frac{1}{2}\theta+ u_{-}(T)\cos\frac{1}{2}\theta$. 
If one assumes that the total derivative term is neglected in the exponential shoulder, no time-derivative terms of the angular variables appear and no monopoles in \eqref{modified state}.

\end{document}